\documentclass[fp]{jpsj3}
\usepackage{txfonts}
\usepackage[svgnames,dvipdfm,rgb]{xcolor}

\title{Reaction Rate Theory of Radiation Exposure and Scaling Hypothesis in Mutation Frequency}

 \author{Yuichiro Manabe$^1$\thanks{manabe{\_}y@see.eng.osaka-u.ac.jp},
 Issei Nakamura$^2$\thanks{nakamura@ciac.ac.cn}, and Masako Bando$^3$\thanks{bando@yukawa.kyoto-u.ac.jp}}
\inst{
$^1$Division of Sustainable Energy and Environmental Engineering,\\ 
Graduate School of Engineering, Osaka University, \\
2-1 Yamada-oka, Suita-shi, Osaka, 565-0871, Japan\\
$^2$State Key Laboratory of Polymer Physics and Chemistry, Changchun Institute of Applied Chemistry, Chinese Academy of Sciences, Changchun, 130022, P. R. China \\
$^3$Yukawa Institute for Theoretical Physics, Kyoto University,\\
Kitashirakawa oiwake-cho, Sakyo-ku, Kyoto-shi, Kyoto, 606-8502, Japan} 

\abst{We develop a kinetic reaction model for cells having irradiated DNA molecules due to ionizing radiation exposure. 
Our theory simultaneously accounts for the time-dependent reactions of the DNA damage, the DNA mutation, the DNA repair, and the proliferation and apoptosis of cells in a tissue with a minimal set of model parameters. 
In contrast to existing theories for radiation exposition, we do not assume the relationships between the total dose and the induced mutation frequency.  
Our theory provides  a universal scaling function that reasonably explains the mega-mouse experiments in Ref.[W. L. Russell and E. M. Kelly, Proc. Natl. Acad. Sci. USA. {\bf 79} (1982) 542.] with different dose rates. 
Furthermore, we have estimated the effective dose rate, 
which is biologically equivalent to the ionizing effects other than those caused by artificial irradiation. 
This value is $ 1.11 \times 10^{-3}  ~\rm{[Gy/hr]}$, which is significantly larger than the effect caused by natural background radiation.}

\begin{document}
\maketitle

\section{Introduction}

Ionizing and non-ionizing radiation exposures are a phenomenon that all living organisms cannot avoid. 
While non-ionizing radiation refers to electromagnetic radiation that does not carry kinetic energy enough to liberate electrons from atoms or molecules, ionizing radiation naturally or artificially generated through nuclear reactions can break molecular bonds and produce harmful free radicals of solvents. These chemical reactions may cause significant physical damage to DNA molecules that encode their genomes in living cells. 
For example, the overwhelming contribution of cellular DNA damages in an aqueous solution is caused by hydroxyl radicals arising from the surrounding water molecules \cite{sonntag}.

Of particular interest in radiobiology over the past decades is genetic mutation induced by irradiation that changes nucleotide sequences of the genome and hence increases the risk of cancers. 
Muller first studied genetic effects of X-rays on Drosophila \cite{muller} and discovered that artificial ionizing radiation gives rise to the mutation. 
The subsequent argument then led to the {\it linear no threshold} (LNT) hypothesis that the carcinogenic risk caused by biological damage due to the ionizing radiation becomes zero at the y-intercept with no artificial radiation exposure. 
That is, there is no safety threshold for radiation exposure. 
Russell and Kelly further examined the mutation frequency by studying the frequency of transmitted specific locus mutations induced in mouse spermatogonial stem cells \cite{russell}. 
Their striking result was that the mutation frequency linearly varies with the total dose of the ionizing radiation within experimental errors, 
whereas their fitting required two different slopes for chronic and acute dose rates. 
Since these studies, a vast amount of literature has emerged on the subject of radiation exposure and genetic mutation. 
Specifically, the deviation of the mutation frequency from the linear slope with the total dose is a matter of controversy \cite{tuibiana}. 

In this paper, we develop a theory for radiation exposure that accounts for the kinetic reaction of irradiated DNA molecules. 
While the study of the molecular dynamics simulation reveals the reaction pathway of the single and double strand breaking of DNA molecules for picoseconds caused by free hydroxyl radicals due to the ionizing radiation \cite{abolfath11}, 
we need a reaction theory for longer time scales from hours to days to consider DNA mutation in cell cycles. 
Our theory shows that the mutation frequency varies with time 
because of the irradiation and the environmental stimuli to DNA molecules. 
This is in reference to counteracting effects among the DNA damage and mutation, DNA repair as well as the proliferation and apoptosis of cells.
In our theory, the key ingredient is the {\it dose rate} that controls the reaction of the system, 
without invoking the total dose $D$ in the theoretical framework.
We show that the observed dependence of the dose rate on DNA mutation frequency in mouse spermatogonial stem cells that cannot be explained by the classical theories falls on the universal scaling function for the low dose rate

\section{Review of Theory}
On the theoretical side, the target theory has been developed originally by Lea\cite{lea}, 
in which individual quanta, or photons, of radiation assumed to be absorbed at sensitive points (targets) in a cell;
They start with the differential equation according to the stimulus-response procedure, 
\begin{eqnarray}
\frac{dN_n}{N_n} =-\frac{dD}{D_0}.
\label{leaeq}
\end{eqnarray}
Here, $N_n$ is the number of normal cells that change to mutated cells with the rate proportional to the total dose $D$. $D_0$ is the unit dose to produce one active event.
The solution of the above equation is given as, 
\begin{eqnarray}
N_n = N_n^0 e^{-\frac{D}{D_0}}, \ \ \ \  N_m = N_n^0-N_n=N_n^0 (1-e^{-\frac{D}{D_0}}),
\label{sollea}
\end{eqnarray}
where $N_m$ is the number of mutated cells, $N_n^0$ is the number of the normal cells before the irradiation.  
Later Chadwick and Leenhouts {\cite{chadwick}} proposed the following formula, by taking into account of the effect of DNA repiar, 
\begin{eqnarray}
N_m = \sigma N_n^0 (1-e^{-\frac{D}{D_0}}),
\label{chadwick1}
\end{eqnarray}
where $\sigma$ is the proportion of the mutated cells that are not repaired.
It should be noted that in the low dose region  ($D \ll D_0$), Eq.\,(\ref{chadwick1}) can be expanded into a linear function of $D$ as 
$N_m \sim \sigma N_n^0 D/D_0$. Thus, the LNT hypothesis is rationalized from the target theory, whereas the dependence of both total dose $D$ and dose rate on the mutation frequency still lies out of this theoretical framework.
Further, to account for the observed deviation from the classical target theory at high dose rate, 
Eq.\,(\ref{chadwick1}) was modified by adding the quadratic term O($D^2$) to the exponent of the exponential function \cite{kellerer}. 
\begin{eqnarray}
N_m = \sigma N_n^0 (1-e^{-\frac{D}{D_1}-\frac{D^2}{D_2}}),
\label{chadwick2}
\end{eqnarray}

\section{Reaction Rate Theory}
\begin{figure}
\includegraphics[width=0.9\columnwidth]{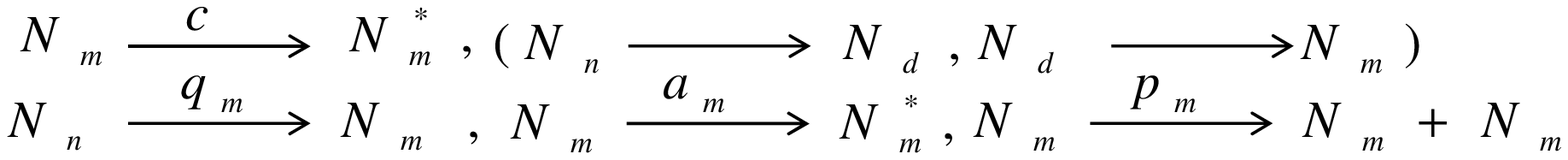}
\caption{Schematic descriptions of the kinetic reaction of cells having mutated DNA molecules. 
We classify the cells in the tissue into two categories, normal and mutated cells. 
We skip an intermediate damaged-cell stage and focus on the relation between normal and mutated cells by expressing it as a rounded-up black-box.}.
\label{fig_full_scheme}
\end{figure}

We now consider a tissue consisting of $N_n(t)$ cells having normal DNA molecules, and $N_m(t)$ cells having DNA mutation. $N_{\rm{max}}$ denotes the maximum number of the cells in the tissue.  At $t = 0$, the tissue is artificially irradiated with the dose rate $d(t)$ [Gy/h]. The total dose $D$ of artificial radiation during the time $t$ is thus given by $D = \int dt d(t)$. In general, cells experience proliferation and apoptosis that are parts of processes of cell reproduction and programmed cell death, respectively. DNA molecules in cells are also damaged through regular biological processes such as cell cycle and environmental irradiation.  The DNA repair process typically responds to the damage in the DNA structure. When the repair of the lesions fails, the DNA mutations can occur. These damage rates may depend on the manner in which the cells are exposed to radiation arising from their surroundings  or in the way they experience metabolism and hydrolysis. In this paper, however, we do not specify the details of such  biological reactions because we do not wish to include a variety of rate constants that cannot be determined or have large uncertainty. Instead, we write the averaged, effective rate of the DNA mutation due to all these relevant natural reactions in time-independent form. Further all living organisms always receive various kinds of stimulus from their surroundings, which cause mutation. We refer these effects to "spontaneous mutation", which should be balanced to their preventive effects. Thus, we introduce $d_{\rm{eff}}$ assigned to the effective dose rate which is biologically equivalent to the ionizing effects other than those caused by artificial irradiation. It should also be noted that $d_{\rm{eff}}$ also includes the effect of natural background radiation. Thus we write the total dose rate $d_{tot}(t)$ in the form of    
\begin{eqnarray}
d_{tot}(t) = d(t) + d_{eff},
\label{totdoserate}
\end{eqnarray}
where $d(t)$ is the dose rate due to artificial irradiation. The kinetic reactions for DNA damage and mutation are schematically shown in Fig.\ref{fig_full_scheme}. 
 
The reaction equations for the numbers of normal cells and cells having mutated DNA molecules, $N_n(t)$ and $N_m(t)$ 
are then written in the form\cite{manabe2012,manabe2013}, 
\begin{eqnarray}
\label{e:full_eq}
\frac{d N_n(t)}{dt}&=& f[N_n(t), N_m(t), d_{\rm eff}, d(t)] \nonumber\\
\dfrac{dN_m(t)}{dt} &=& c [d_{tot}(t)] N_n(t)-\bigl( q_m[d_{tot}(t)]+a_m-p_m \bigr) N_m(t).
\end{eqnarray}
Here, the the reaction rates with the subscripts $n$ and $m$ indicate that they correspond to normal and mutated cells, respectively. $c[ d_{tot}(t)]$ is the reaction rates for the mutation of damaged DNA molecules. 
$q_m[ d_{tot}(t)]$ are the reaction rates for the death of mutated cells.
They are caused by stimulus, and depend on $d_{tot}$.
The parameters,  $p_m$ and $a_m$, denote the reaction rates for the proliferation and the apoptosis of mutated cells, respectively.
They are independent of the stimulus $d_{tot}$. 
Those parameters must be rationalized by more microscopic models that account for mitosis, 
DNA mutation and repair, metabolism, and all other reactions in cell cycles. 
Notice that a generic theory must account for the chemical reaction between the DNA damage and repair. 
Here, however, we do not assume any reaction model. 
Just we assume their general characteristic features. As for the reaction rates, $c[d_{tot}(t)], q_m[d_{tot}(t)]$, the cells respond to stimulus; 
the energy $\Delta \epsilon$ deposited in a single cell having the mass $\rho$ [kg] during the short time $\Delta t$ as 
\begin{eqnarray}
\Delta \epsilon =\rho  d_{tot} \Delta t.
\label{enerrgye}
\end{eqnarray}
We now introduce $P_n (\Delta \epsilon) $ and $P_m (\Delta \epsilon) $, the probability that DNA mutation in a normal cell and the death of a cell having DNA mutation occur, respectively.
The nature of these probabilities must be rationalized by more microscopic models that account for mitosis, DNA mutation and repair, metabolism, and all other reactions in the cell cycle.
In the case of low dose rates, we simply write these probabilities proportional to the energy in the form of  $P_n (\Delta \epsilon) =p_n \rho d_{tot} \Delta t$ and $P_m (\Delta \epsilon) = p_m \rho d_{tot} \Delta t $. 
Thus, the number of normal cell whose DNA molecules are mutated during $\Delta t$ is given by
\begin{eqnarray}
\Delta N_m &=& P_n (\Delta \epsilon) N_n = p_n \rho d_{tot} \Delta t N_n, \nonumber\\
\frac{\Delta N_m}{\Delta t} &=& p_n \rho d_{tot} N_n.  
\label{cqepsironn}
\end{eqnarray}

Similarly the number of mutated cell which dies during $\Delta t$ is given by 

\begin{eqnarray}
\Delta N_m &=& -P_m (\Delta \epsilon) N_m = -p_m \rho d_{tot} \Delta t N_m, \nonumber\\
\frac{\Delta N_m}{\Delta t} &=& -p_m \rho d_{tot} N_m.  
\label{cqepsironm}
\end{eqnarray}

Finally we express\cite{comment}  
\begin{eqnarray}
\label{e:ffull_eq}
\dfrac{d N_m(t)}{dt} = p_n \rho d_{tot} N_n(t)-[p_m \rho d_{tot}(t) +a_m-p_m]N_m(t).
\end{eqnarray}
Therefore
\begin{eqnarray}
c[d_{tot}]&=&p_n \rho d_{tot}(t) \equiv c  d_{tot}(t) , \nonumber\\
q_m[d_{tot}(t)]&=&p_m \rho d_{tot}(t) \equiv q_m  d_{tot}(t) ,  
\label{cqepsiron}
\end{eqnarray}
where we have introduced the constants $c$ and $q_m$, which denote the first coefficients of Taylor series for $c(\epsilon)$ and $q_m(\epsilon)$ at $\epsilon = 0$.  
Thus, Eq.\,(\ref{e:full_eq}) for $N_m (t)$can be cast into the following form with the time-independent reaction rates,
\begin{eqnarray}
\label{e:ffull_eq}
\dfrac{d N_m(t)}{dt}&=& c d_{tot}(t) N_n(t)-[q_m d_{tot}(t) +a_m-p_m]N_m(t), 
\end{eqnarray}
where the parameters are $c, q_m$ and $a_m-p_m$.

\section{Mutation Frequency}
In this section, we apply the reaction rate theory,  Eq.\,(\ref{e:ffull_eq}), to the case of the mutation frequency obtained by the 
pronounced experiment for mega-mouse projects \cite{russell}. 
In this case, the model tissue primarily consists of almost normal cells only, 
we take the approximation,  $N_n(t) \sim N_{\rm{max}}$ during the reaction. 
Our rationale for this treatment is that the experimental data of the mutation in mouse spermatogonial stem cells indicate that
 $N_m(t)/N_n(t)$ is close to $10^{-5}$. 
Thus, the number of the normal cells whose DNA molecules are mutated during $\Delta t$ is given by
 $c \epsilon  N_n(t) \sim c \epsilon N_{max}$.

Replacing $N_n (t) $ by $ N_{\rm{max}} $ of the first term of $N_m$  in Eq.\,(\ref{e:ffull_eq}), we obtain
\begin{eqnarray}
\label{e:decoupled}
\frac{d N_m(t)}{dt} &=& cd_{tot}(t) N_{\rm{max}}-[q_m d_{tot}(t) +a_m-p_m]N_m(t).
\end{eqnarray}
We note that the differential equation of the number of mutated cells is thus decoupled from the number of normal cells.
Because our primary interest is the mutation frequency, we consider
\begin{eqnarray}
\label{defF}
F(t)=\frac{N_m(t)}{N_n(t)} \sim \frac{N_m(t)}{N_{max}}.
\end{eqnarray}  
Eq.\,(\ref{e:decoupled}) then follows   
\begin{eqnarray}
\label{difeqF}
\frac{d F(t)}{dt}=\gamma d_{tot}(t) -[ \beta d_{tot}(t) + \mu)]F(t), 
\end{eqnarray}
where we have symbolically written three parameters  as,  
\begin{eqnarray}
\label{parametersinF}
\gamma=c(\rho), \ \  \beta&=&q_m(\rho),\ \  \mu = a_m - p_m, \nonumber\\
           d(t)_{tot}&=&d(t)+d_{\rm{eff}},  
\end{eqnarray}    
with  $\rho = M/N_{\max}$ denoting the average weight density of the cells in the tissue, $M$ denoting the total mass of living object.
 
\begin{figure}
\includegraphics{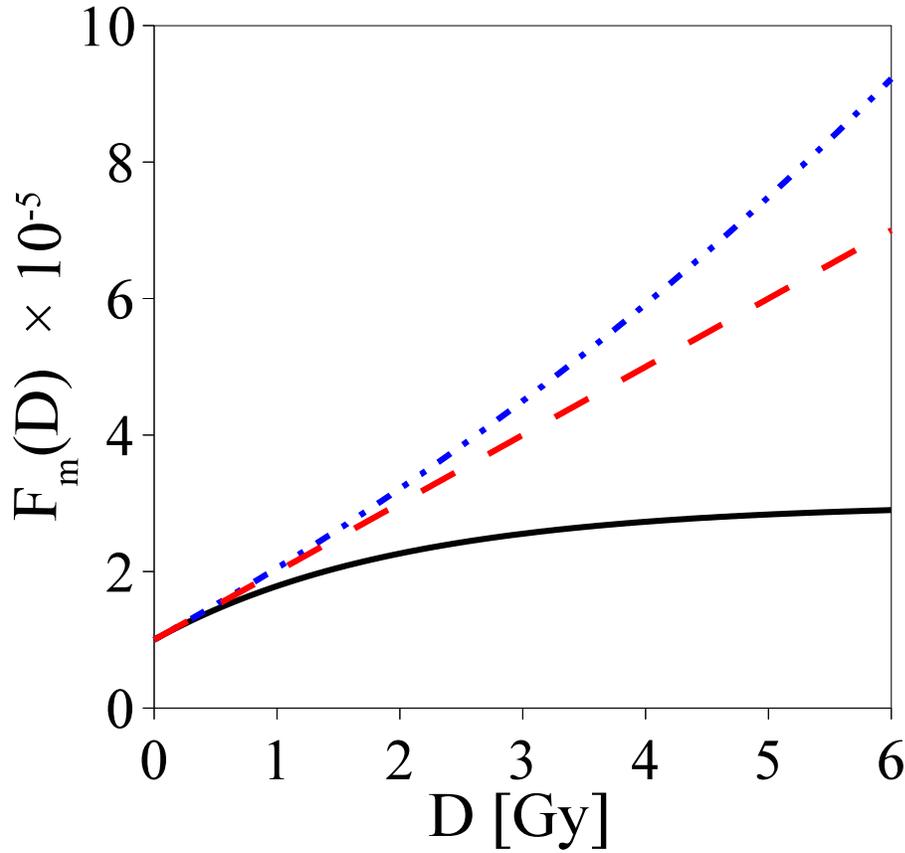} 
\caption{$F_m(D)$ vs. $D$ for the condition of values of $[c_m d_{tot} + \mu]$. (1)$[c_m d_{tot} + \mu] > 0$ (solid line) (2)$[c_m d_{tot} + \mu] = 0$ ( dashed line) (3) $[c_m d_{tot} + \mu] < 0$ (dot-dashed line)}
\label{F_m_D} 
\end{figure}

For conciseness, we consider  $d(t) = d \theta (t)$, where $\theta (t)$ is the step function. 
Thus, we have 
\begin{eqnarray}
\label{eqFtcostrate}
d_{tot} &=& d_{\rm{eff}} \,\, {\rm for} \,\, t< 0, \nonumber\\
d_{tot} &=& d_{\rm{eff}}+d \,\, {\rm for} \,\, t \geq 0.
\end{eqnarray} 
For the solutions of Eq.\,(\ref{difeqF}), we have three classifications with respect to $\ \beta d_{tot} + \mu >0\ $, $\ \beta d_{tot}+ \mu = 0\ $, 
and $\ \beta  d_{tot}+ \mu < 0\ $. In Fig.\,\ref{F_m_D}, we have illustrated the possible scenarios with respect to the cases with $[\beta d_{tot}+ \mu]$. When the specific condition with $\ \beta d_{tot}+ \mu = 0\ $ is satisfied, the solution of Eq.\,(\ref{eqFtcostrate}) becomes,  
\begin{eqnarray}
\label{eqFsolzero}
F(t) &=&\gamma d_{tot} t+ F(0). 
\end{eqnarray}

If we stop the constant irradiation at $t=T$ , which amounts the total dose,
which is total amount of artificial irradiation, 
$D=dT$,
and the mutation frequency just after irradiation becomes,  
\begin{eqnarray}
\label{eqFsolzeroD}
F(D) &=& \gamma  d_{tot}  T + F(0) =\gamma \biggr( 1 + \frac{d_{\rm{eff}}}{d} \biggl) D+F(0).
\end{eqnarray}
Thus, the LNT hypothesis remains intact with any value of $D$. 
Note that this holds only if $ \beta d_{tot} + \mu=0$ is exactly satisfied. 
Certainly, this strict condition is unlikely to occur in living organisms and difficult to be externally controlled. 
We do not consider the case $ \beta d_{tot} + \mu < 0$ as we we would like to consider the case when $F(D)$ have finite value in the limit of $D \to \infty$.

When $\beta d_{tot} + \mu > 0$, the solution of  Eq.\,(\ref{eqFtcostrate}) can be analytically solved;
\begin{eqnarray}
\label{eq_solneq}
F(t)=\biggl( \frac{\gamma d_{tot} }{\beta d_{tot}  + \mu} - F(0) \biggr) [1 - e^{-(\beta d_{tot}  + \mu)t}] + F(0).
\end{eqnarray}
Again, let us express $F$ in terms of the total dose $D$ when the constant irradiation stops at $t=T=D/d$ 
because $D$ and $d$ are more common setup in the experiments.
\begin{eqnarray}
\label{eq_sol}
F(D) &=& \biggl\{\dfrac{\gamma d_{tot} }{\beta d_{tot}  + \mu } - F(0)\biggr\} \nonumber\\
&\times& \biggl(1 - e^{-( \beta d_{tot}  + \mu ) D/d} \biggr) + F(0).
\end{eqnarray}
For  the small total dose with $D \ll d/ [ \beta (d + d_{\rm{eff}}) + \mu]$, 
Eq.\,(\ref{eq_sol}) is expanded to 

\begin{eqnarray}
\label{eq_exp}
F(D) & \to & \biggl\{\dfrac{\gamma d_{tot} }{\beta d_{tot}  + \mu } - F(0)\biggr\} \nonumber\\
&\times& \biggl( ( \beta d_{tot}  + \mu ) /d \biggr) D + F(0) +O(D^2). 
\end{eqnarray}

Thus, the LNT hypothesis holds only with the small total dose whose condition depends on the dose rate $d$. 

The steady state solution of Eq.\,(\ref{eqFtcostrate}),  
$\bar F$ is derived by setting $\frac{d F(t)}{dt}= 0$ and becomes 
\begin{eqnarray}
\label{eq_solf0eff}
\bar F = \frac{ \gamma d_{tot}}{\beta d_{tot} + \mu} ,
\end{eqnarray}
which corresponds to the asymptotic value as $t$ goes to infinity. 
Note that we have already assumed that $\beta d_{tot} + \mu > 0$,
otherwise  $\bar F \to \infty $.

We then identify the control $F(0)$ with the steady state solution without the artificial radiation (i.e., $d = 0$),  
\begin{eqnarray}
\label{eq_initial}
F(0)=\frac{\gamma d_{\rm{eff}}}{\beta d_{\rm{eff}} + \mu}.   
\end{eqnarray}

In terms of the above steady state expressions of $\bar F, F(0)$,  Eq.\,(\ref{eq_solneq}) is expressed as, 
\begin{eqnarray}
\label{eq_solneq2}
F(t)=[\bar F - F(0)] [1 - e^{-(\beta d_{tot}  + \mu)t}] + F(0).
\end{eqnarray}

\section{Numerical Results}

In principle, the three parameters, $\gamma \ $, $\beta \ $, and $\mu \ $ should be derived from more microscopic models that account for the relevant phenomena such as cell cycles, breaking DNA strands and base pairs due to irradiation and the chemical reactions in cell cycles, and repairing them. 
In our kinetic reaction model, however, they are the model parameters determined by experimental data. 
Using Eqs.\,(\ref{eq_sol}) and the pronounced experimental data in Ref.\,\cite{russell} for mouse spermatogonial stem cells, we have determined 
\begin{eqnarray}
\label{parameters}
\gamma &=& 2.91 \times 10^{-5} \,\, {\rm [1/Gy]},\nonumber\\ 
\beta &=& 1.00 \times 10^{-1} \,\, {\rm [1/Gy]}, \nonumber\\ 
\mu &=& 3.13 \times 10^{-3} \,\,    {\rm [1/hr]}, 
\end{eqnarray}
by the least squares fitting\cite{comment_2}. 
Eq.\,(\ref{eq_initial}) then leads to
\begin{eqnarray}
\label{eq_m_eff}
d_{\rm{eff}}  &=& \dfrac{F(0)\mu}{\gamma - F(0)\beta } = 1.11 \times 10^{-3} ~\rm{[Gy/hr]}.
\end{eqnarray}
We note that this value is significantly larger than $2.74\times 10^{-7}$ [Gy/hr] due to the natural background radiation. 
Thus, our result indicates that the natural damage of the DNA molecules arises primarily from 
the stimulus other than natural background radiation, and it may comes from the chemical reactions in cell cycles. 
Note also that the value in Eq.\,(\ref{eq_m_eff}) is of the same order as the one for humans (8.4 mGy/hr) due to the double-strand DNA breaks caused by endogenous reactive oxygen species \cite{tuibiana}.


Our model for mouse spermatogonial stem cells shows that the mutation frequency becomes twice as the control due to the spontaneous mutation
when the total dose $D$ for a year reaches $\sim 10$ [Gy]. 
This value  shares a similar feature as the 
so-called 'doubling dose,' the standard concept in radiation biology, 
and is suggested to be surprisingly similar values among humans, mice, and drosophila: 
in the range of $0.1-10.00$ [Gy] \cite{neel} where our result falls on as well. 
This similarity, together with our result, may also imply that they commonly receive the risk of the spontaneous mutation per gene.
However, our kinetic modeling indicates that while the doubling dose is a widely accepted concept in radiation biology, 
the total dose $D$ is not a fundamental measure to account for the mutation frequency.
We welcome further experiments to clarify the new concept based on the dose rate $d(t)$. 

\section{Scaling}
\begin{figure}
\includegraphics{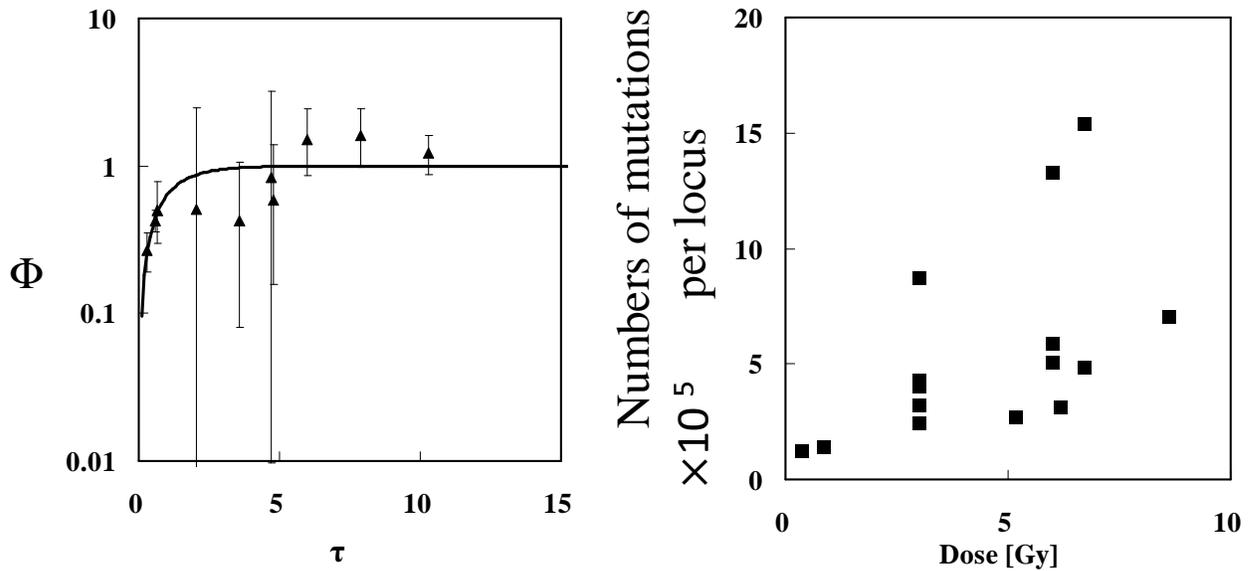} 
\caption{Scaling function $\Phi$ vs. scaled time $\tau$. Solid line and triangular points with errors indicate theory and experiment \cite{russell}, respectively. For comparison, we reproduced the original experimental data  from Ref.\,\cite{russell} in the inset.: The x-axis and y-axis denote the time of exposure [hr] and mutation frequency $\!$ $\times$ $10^5$ per locus. Note that each data point has the different  value of the dose rate $d$.}
\label{fig_mfrq} 
\end{figure}

We now cast Eq.\,(\ref{eq_sol}) into
\begin{eqnarray}
\label{eq_scaling}
&\Phi \equiv \dfrac{F[\tau] - F(0)}{\overline{F} - F(0)]} = 1 - \exp{[-\tau]},  \nonumber\\
&\ \tau = [\beta (d + d_{\rm{eff}}) + \mu] t,
\end{eqnarray}  
where we have introduced the scaled time $\tau $. Importantly, Eq.\,(\ref{eq_scaling}) indicates that, in general, mutation frequencies with the low-dose irradiation fall on the universal scaling function $\Phi$. 
To illustrate our scaling function, we have used the same experimental data  \cite{russell} for fitting our model parameters (Fig.\,\ref{fig_mfrq}). 
The inset of the figure shows that the original data points scatter in the range of 0.056-7143 [hr], in contrast to the scaled ones. 
We note that each experimental data point corresponds to cases of {\it different} values of the dose rate $d$; 
our fitting was performed with respect to not a single value of the dose rate, but multiple values of the dose rate.
Thus, our theory shows qualitative agreement with the experimental data without classifying the dose rate $d$.

\section{Conclusion}

In summary, we performed a kinetic reaction modeling for the proliferation and apoptosis of cells,
DNA damage and mutation due to the environmental stimuli and irradiation, as well as DNA repair. 
%
Our theory's key features are that our kinetic rate equations include the dose rate $d(t)$ in the rate constants. 
In addition, the rate equations for normal cells and cells having damaged DNA molecules due to the low dose irradiation are decoupled from that for cells having mutated DNA molecules. 
Despite the simplicity of the equations, we are able to qualitatively explain unaccountable behavior in the pronounced experiment for mega-mouse projects \cite{russell} in which two linear slopes for the mutation frequency vs. the total dose rate exist with respect to acute and chronic irradiations. 
Thus, our theory suggests that the total dose $D$ is not a fundamental measure to study irradiation as deduced from the systematic relationships for the dose rate vs. the induced mutation frequency \cite{vilenchik00,vilenchik06}. 
Depending on the rate constants, the number of the cells having mutated DNA molecules may continue to monotonically increase. 
While no one desires this phenomenon clinically, this article demonstrates a lesson on the importance of accurate control of dose rate in the study of mutation frequencies and presumably, cancer risks. 
Importantly, our theory predicts that all the experimental data of mouse spermatogonial stem cells with low-dose rates fall on the universal scaling function $\Phi$ with the scaled time $\tau$ [Fig.\,{\ref{fig_mfrq}}].
Because this experiment necessitated seven million mice for sampling, similar data for mice cannot be obtained by the current code of ethics in experiments. Thus, validating the universality in different species would be challenging as well as interesting to us as a future study.










\begin{thebibliography}{20}
\bibitem{comment}%
  {In the limit of very high dose rate, we expect that the DNA molecules are
  always damaged. Thus, these probabilities should be unity per unit time,
  regardless the deposited energy.}
\bibitem{comment_2}%
  {In Ref.\cite{russell},
  we found that one data point does not appear to be classified by the chronic
  and acute radiations that the authors defined. While we have eliminated this
  data point, there is no bias or prejudice in our selecting the data for
  fitting.}
\end{thebibliography}
\end{document}